\documentclass[cameraready]{Interspeech}
\usepackage[most]{tcolorbox}
\usepackage{xcolor}
\usepackage[sort,compress]{cite}
\usepackage{CJKutf8}
\title{Edge–Cloud Collaborative Speech Emotion Captioning via Token-Level Speculative Decoding in Audio-Language Models}

\author[affiliation={1}, orcid=0009-0002-3098-9692]{Xiangyuan}{Xue}
\author[affiliation={2}, orcid=0009-0003-7385-7136]{Jiajun}{Lu}
\author[affiliation={3}, orcid=0000-0001-7922-9788]{Yan}{Gao}
\author[affiliation={4}, orcid=0000-0002-6825-7473]{Gongping}{Huang}
\author[affiliation={2}, orcid=0000-0003-3806-1493]{Ting}{Dang}
\author[affiliation={1}, orcid=0000-0002-6047-4158]{Hong}{Jia}
\address{
    $^1$ The University of Auckland, New Zealand \\
    $^2$ The University of Melbourne, Australia  \\
    $^3$ The University of Cambridge, United Kingdom \\
    $^4$ Wuhan University, China
}

\email{xxue752@aucklanduni.ac.nz}


\usepackage{comment}
\usepackage{soul, xcolor}

\usepackage{placeins} % provides \FloatBarrier
\usepackage{booktabs,tabularx}
\begin{document}

\maketitle

\begin{abstract}
% Speech Emotion Captioning (SEC) uses large audio–language models to generate rich, context-aware descriptions of affect from speech, but deployment is constrained by autoregressive decoding cost, limited edge compute, and privacy concerns around biometric audio. Small audio–language models make on-device SEC feasible, yet their reduced capacity destabilizes subtle paralinguistic cues and weakens fine-grained affective grounding. We propose Uncertainty-Guided Speculative Decoding (UGSD), an entropy-driven edge–cloud framework in which a lightweight edge model drafts captions and selectively escalates only high-uncertainty token blocks to a stronger cloud verifier, without transmitting raw waveforms. This token-level draft-and-verify design improves caption fidelity and efficiency while reducing communication. On MER2024, UGSD achieves 21.6\% and 49.3\% relative gains in BLEU-1 and BLEU-4 over the edge-only baseline on English, and 21.0\% and 62.7\% gains on Chinese, while cutting end-to-end latency from 40.21 s to 28.67 s (1.40×) and increasing output token throughput from 1.53 to 13.05 tokens/s. Ablations confirm the benefits of uncertainty-triggered collaboration and analyze how draft block size and verifier capacity shape the quality–efficiency–privacy trade-off.
Speech Emotion Captioning (SEC) leverages large audio–language models to generate rich, context-aware affective descriptions from speech. However, real-world deployment remains challenging due to the substantial computational demands on resource-constrained edge devices and the privacy risks of transmitting biometric audio. While smaller audio–language models enable efficient on-device SEC, their limited capacity often weakens subtle paralinguistic modeling and fine-grained affective grounding. We propose an edge–cloud collaborative framework based on \textbf{U}ncertainty-\textbf{G}uided \textbf{S}peculative \textbf{D}ecoding (\textbf{UGSD}). A lightweight edge model drafts captions locally, and only high-uncertainty token blocks are selectively escalated to a stronger cloud verifier for validation. Experiments on the MER2024 benchmark demonstrate substantial BLEU improvements up to 62.7\%. UGSD further achieves 1.4× lower latency and 8.5× higher token throughput compared to an edge-only model. These results empirically characterize the quality–efficiency–privacy trade-off in deployable SEC systems.\looseness=-1

% Speech emotion captioning describes affect from speech, but deployment is limited by on-device compute and privacy constraints. We propose an uncertainty-aware edge-to-cloud collaborative decoding framework with token-level speculative decoding: an on-device audio-language model drafts captions, and a cloud verifier is invoked only when predictive uncertainty exceeds a threshold, transmitting only compact information. On MER2024, we improve over the on-device baseline (Edge) by 21.6\% BLEU-1 and 49.3\% BLEU-4, while reducing end-to-end time from 40.21\,s to 28.67\,s (1.40$\times$) and increasing throughput from 1.53 to 13.05 tokens/s. Similar gains are observed on the Chinese subset (+21.0\% BLEU-1, +62.7\% BLEU-4). Ablations confirm the benefits of uncertainty-driven collaboration and analyze the effects of draft block size and verifier capacity.
\end{abstract}

\section{Introduction}

Speech Emotion Recognition (SER) is undergoing a paradigm shift from discrete categorical classification toward Speech Emotion Captioning (SEC)~\cite{xu2024secap,liang2024aligncap,wang2025bridging}. Unlike traditional models that output isolated labels like ``Happy'' or ``Sad,'' SEC leverages Large Audio-Language Models (LALMs) to generate rich, context-aware descriptions of affective states from speech signals. By synergizing high-capacity audio encoders, such as HuBERT~\cite{hsu2021hubert} or Whisper~\cite{radford2023whisper}, with the reasoning prowess of Large Language Models (LLMs), LALMs can articulate subtle paralinguistic nuances, such as a ``tremulous undertone of anxiety'' or ``sarcastic inflection''.  These capabilities are foundational for the next generation of empathetic AI assistants and real-time accessibility tools.

Despite their descriptive superiority, deploying these foundation models on edge devices is severely constrained by model scale and decoding cost. The massive memory footprint of 7B+ parameter LALMs typically exceeds the RAM budgets of commodity edge hardware, and their autoregressive token-by-token generation leads to substantial inference latency. Consequently, state-of-the-art emotion captioning remains tethered to cloud-dependent architectures%\td{here the reference should be cloud based but you cite cloud-edge collabraitve?}
, introducing significant communication overhead and raising critical privacy concerns regarding the transmission of sensitive biometric audio data.

Recent work has increasingly explored Small Audio Language Models (SALMs)~\cite{qwen2025qwen25,su2025audio,choi2022temporal} via compression techniques~\cite{han2016deepcompression,lin2024awq,xu2021mixed} or knowledge distillation techniques~\cite{hinton2015distilling,wang2025cross,yang2026attention} to enable on-device speech understanding tasks such as automatic speech recognition and emotion recognition. However, a recurring limitation of compact models remains: as model capacity decreases, the capturing of subtle paralinguistic representations may deteriorate~\cite{qwen25omni},
% \td{here the references should be the ones that highlight slms limitations in this; you can cite papers that shows slms worse than llms in speech tasks if any}
and empirical evaluations on speech tasks show that larger pretrained models often outperform smaller variants~\cite{zaiem2023speech}, potentially limiting the mapping from fine-grained acoustic evidence to target semantic caption. This limitation becomes especially pronounced in SEC, where the task demands fine-grained modeling of co-occurring and subtle emotional cues, as well as their faithful grounding in acoustic evidence to generate semantically rich, context-aware natural language descriptions. %Compact on-device models often struggle to preserve such affective granularity.

% enable on-device speech understanding such as speech recogntion, emotion recogntion \td{here you can cite qwen small version paper}. However, we observe a recurring limitation of compact models in general: as model capacity decreases, the representation of subtle paralinguistic cues becomes less stable, and the mapping from acoustic evidence to fine grained affective language or paralinguistic cues weakens\td{}, limiting the model ablity to interpete nuanced emotion interpretations grounded in the audio, posibly limiting the captioning capbalities.

%In practice, small models tend to over smooth ambiguous or mixed emotion signals and produce generic emotions rather than committing to nuanced interpretations grounded in the audio\td{here you can cite wenda zhang's paper from our lab}. This ``emotional dilution'' limits their ability for emotion understanding when the utterance contains layered affect, irony, or low intensity cues.

Edge–cloud collaborative inference offers a promising middle ground by keeping most computation on the device, which reduces latency and improves privacy, while selectively invoking the cloud to enhance caption reliability.  However, most existing approaches adopt static partitioning strategies across network layers or predefined modules~\cite{shi2019improving,hao2024hybrid,wang2024clouddevice}%\td{double check refereces to be ture..}
. Once the partition is fixed, all tokens follow the same offloading path, regardless of their prediction difficulty. Such designs fail to account for token-level variability during autoregressive generation. This limitation is particularly critical for SEC, where subtle affective cues and fine-grained semantic details are often concentrated in a small subset of emotionally salient tokens. These tokens are inherently more uncertain and sensitive to representation instability. Consequently, treating all tokens equally can either waste computation on easy tokens or under-allocate resources to difficult ones\cite{hao2024hybrid,wang2024clouddevice}%\td{not the correct references. Please review all references thoroughly to ensure they are cited correctly. Incorrect citations compromise academic integrity. }
, potentially leading to degraded caption fidelity and affective grounding. Moreover, processing all tokens (or long continuous segments) in the cloud results in unnecessary computational and communication overhead. Consequently, prior work lacks a principled, token-level mechanism that enables efficient and effective edge-cloud collaborative learning for SEC.% can dynamically identify unreliable predictions and selectively refine them during SEC decoding.%\td{are your references correct? Please be careful!!! 5 and 6 were cited for cloud only structures, but now as collaborative learning. [1,2] here also don't correspond to what we aim to claim here.}

% \begin{figure*}[t]
%   \centering
%   \includegraphics[
%     width=\textwidth,
%     height=0.30\textheight,
%     keepaspectratio
%   ]{p11.pdf}
%   \caption{Overview of the uncertainty-aware collaborative decoding framework for Speech Emotion Captioning.}
%   \label{fig:overview}
% \end{figure*}

\begin{figure*}[t]
  \centering
  \includegraphics[width=1\textwidth]{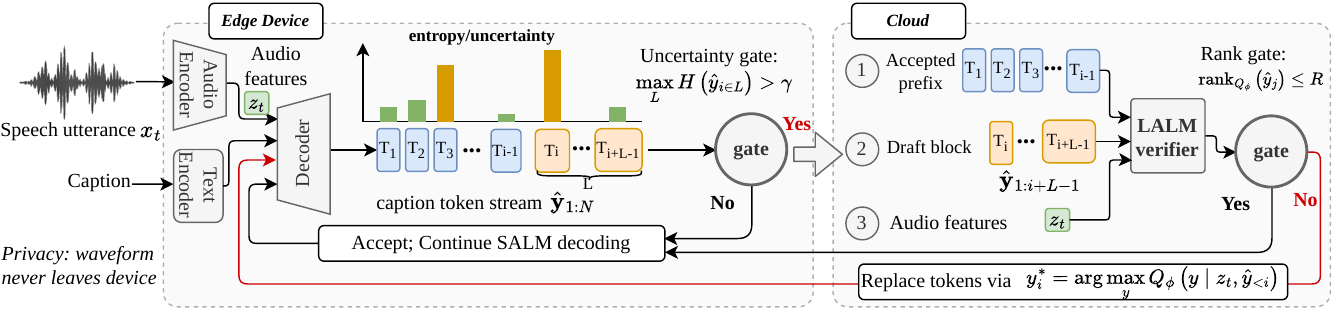}
  \caption{Overview of the uncertainty-aware collaborative decoding framework for Speech Emotion Captioning.}
 \label{fig:overview}
\end{figure*}

To address this gap, we propose \textbf{Uncertainty-Guided Speculative Decoding} (\textbf{UGSD}), an uncertainty-aware edge-cloud collaborative framework that uses token-level entropy as a proxy for semantic unreliability. During edge decoding, tokens with high predictive entropy are selectively sent to a cloud verifier for validation and refinement, while confident tokens remain local, enabling fine-grained adaptive offloading. To our knowledge, this is the first entropy-driven, token-level adaptive collaboration framework tailored for speech emotion captioning. %\hj{I think we used two datasets? So in here we say: To evaluate our framework, we conducted on representative datasets including MER2024 and xxx.}\td{one} 
Experiments conducted on MER2024 dataset across two different languages demonstrate that UGSD improves significantly over the edge-only baseline, with 21.6\% to 76.4\% relative improvements in terms of BLEU-1, BLEU-4, METEOR, and ROUGE-L respectively. It further reduces end-to-end inference time from 40.21\,s to 28.67\,s (1.40$\times$ faster) and increasing output token throughput from 1.53 to 13.05 tokens/s. Moreover, only a small fraction %\td{the precise portion here}
of tokens (18.2\%) require cloud verification, which substantially reduces data transmission and strengthens privacy preservation compared to fully cloud-based inference. This work opens the avenue for effective, efficient, and privacy-preserving edge–cloud decoding for SEC.

\section{Related Work}

\noindent \textbf{Speech Emotion Captioning.} Speech Emotion Captioning (SEC) reframes emotion understanding from categorical recognition to open-ended natural language generation. Compared with conventional recognition, which predicts discrete labels and has been widely studied~\cite{oates2019robust,pepino2021emotion,leem2023computation,halim2025token,jia2026decoding}, SEC aims to describe nuanced affective cues and their expression in speech using descriptive captions~\cite{xu2024secap,liang2024aligncap,wang2025bridging}. 
% \td{13-15 are not captioning..}
Early work such as SECap~\cite{xu2024secap} establishes a speech-to-caption pipeline and demonstrates the feasibility of producing emotion descriptions from speech. More recent efforts improve caption robustness and alignment quality. For example, AlignCap~\cite{liang2024aligncap} explores alignment to human preferences to enhance caption naturalness and consistency. While this line of work clarifies task definitions and training objectives, it primarily emphasizes model design and caption quality with LALMs. Deployment-level constraints, including on-device privacy and end-to-end latency remain underexplored.

\vspace{5pt}
\noindent \textbf{Speculative Decoding for Efficient Generation.}
Speculative decoding mitigates the sequential bottleneck of autoregressive generation via a draft-and-verify strategy~\cite{leviathan2023speculative}. A lightweight draft model proposes a short token block, and a stronger verifier checks the block in a single forward pass, accepting tokens that are consistent with its own predictions and correcting those that are not to preserve output fidelity. Speculative decoding has been recently extended to speech tasks~\cite{lin2025accelerating,okabe2025simultaneous,wei2025specasr}. For example, SpecASR applies draft-and-verify to LLM-based automatic speech recognition and reports substantial latency reductions without sacrificing recognition accuracy in a single-system setup~\cite{wei2025specasr}. However, these methods are limited to single-node speculative decoding, aiming to reduce GPU forward passes on a single device, while failing to account for communication overhead and effective transmission strategies in edge–cloud distributed environments.
% edge–cloud collaboration, token-level uncertainty handling and privacy-preserving communication during generation have not yet been widely examined.\td{jiajun, not clear. what is the gap here?}
% \td{@jiajun, we don't claim centralized acceleration here as it is not related to speculative decoding. Can you let me know if this work i) focuses on large models only but not small models? and ii) does it use uncertain-aware tokens or still passing all tokens to the cloud?}

\vspace{5pt}
\noindent \textbf{Edge-Cloud Collaborative Inference.}
% Edge-Cloud collaborative inference is a promising paradigm for deploying large models efficiently. A lightweight edge model generates initial predictions, while a stronger cloud model performs selective refinement for uncertain or difficult tokens~\cite{hao2024hybrid}. Edge-Cloud collaborative learning further reduces communication overhead by prioritizing uncertain or high-impact information instead of continuously streaming intermediate representations~\cite{wang2024clouddevice}. Despite these advances, there is no prior work exploring edge--cloud collaborative inference tailored to SEC, particularly a token-level uncertainty-aware design that jointly improves generation quality, efficiency, and privacy preservation.
Edge–cloud collaborative inference is a promising paradigm for deploying large models under real-world constraints. By running a lightweight draft model on the device and offloading only selected work to a stronger cloud model, it can reduce on-device computation, amortize heavy inference on the server, and control communication overhead while still improving prediction quality~\cite{hao2024hybrid,wang2024clouddevice}. However, many edge–cloud collaborative inference frameworks are built around layer or module-level partitioning and intermediate-feature transmission between device and cloud~\cite{shi2019improving}. They overlook token-level control for improved latency–accuracy trade-offs.

%\td{not strong, if already done, what is your novelty?} %As a result, the potential of token-level, uncertainty-aware speculative collaboration—jointly targeting caption quality, efficiency, and privacy in SEC—remains largely unexplored.

\section{Edge-Cloud Collaborative Learning}
As shown in Figure~\ref{fig:overview}, the edge-cloud collaborative framework consists of a lightweight draft model on the edge device and a stronger verifier model in the cloud. Our edge–cloud collaborative framework uses a two-stage decoding strategy to improve latency-accuracy trade-offs.

\subsection{Draft Model with SALM}
\vspace{-1mm}
Given a speech utterance $x_t$, the waveform is first encoded by the edge-side draft model (a SALM), which autoregressively generates a token sequence $\hat{y}_{1:N} = \{ \hat{y}_1, \hat{y}_2, \dots, \hat{y}_N \}$ representing a preliminary speech emotion caption.

Unlike speculative decoding approaches that offload all drafted tokens to the cloud for verification~\cite{leviathan2023speculative}, we propose an uncertainty-aware collaborative inference framework. Instead of verifying every drafted token, only high-uncertainty tokens are selectively sent to the cloud for verification and correction, reducing redundant cloud computation.

% To enable uncertainty-aware collaborative inference, the draft model als o estimates token-level predictive uncertainty during decoding. We use entropy as a gating signal to decide whether a drafted token block needs to be checked by the cloud model. Cloud verification is triggered only when the draft is uncertain rather than being applied to every block by default. This keeps the number of cloud interactions under control, avoids unnecessary verification steps, and still allows the model to correct parts that are less reliable.
Specifically, the uncertainty of each generated token $i \in [1,N]$ is measured by entropy:
\vspace{-5pt}
\begin{equation}
\scriptsize
H(\hat{y}_i) = - \sum_{k=1}^{|\mathcal{V}|} p_\theta(y_i=k \mid x_t, \hat{y}_{<i}) \log p_\theta(y_i=k \mid x_t, \hat{y}_{<i}),
\end{equation}
\noindent where $\mathcal{V}$ denotes the vocabulary space and $p_\theta$ is the draft model probability distribution.

Tokens with high entropy are considered unreliable. When the maximum uncertainty within a token block of length $L$ exceeds a predefined threshold $\gamma$, i.e.,
\begin{equation}
% \small
% \frac{1}{L} \sum_{i=j}^{j+L-1} H(\hat{y}_i) > \gamma,
\max_{i=j,\dots,j+L-1} H(\hat{y}_i) > \gamma,
\end{equation}
the corresponding length-$L$ token span is transmitted to the cloud-side LALM verifier.

\subsection{Verifier with LALM}
\vspace{-1mm}
The cloud verifier is invoked only when high uncertainty is detected during draft generation. Upon escalation, the cloud model evaluates the drafted tokens and checks whether they are consistent with its own predictions. Tokens that align with the verifier's high-confidence predictions are accepted; otherwise, they are rejected and corrected.

Specifically, the cloud verifier computes token probability distributions using a larger model for all draft positions in a single forward pass~\cite{leviathan2023speculative}. It jointly processes (i) the previously accepted token prefix $\hat{y}_{<i}$, (ii) the $L$ drafted token IDs in the current block, and (iii) the abstract acoustic representation $\bm z_t$ extracted on-device from $x_t$. For privacy preservation, the raw waveform $x_t$ remains on the device and is never transmitted to the cloud.

Formally, let $Q_\phi(y_i \mid \bm z_t, y_{<i})$ denote the token distribution produced by the cloud verifier model (LALM). We adopt a rank-based token acceptance rule: a drafted token $\hat{y}_i$ is accepted if
\begin{equation}
\text{Rank}_{p_\phi}(\hat{y}_i) \leq R,
\end{equation}
where $R$ is a predefined acceptance threshold. In other words, a token is accepted if it falls within the top-$R$ most probable tokens under cloud $Q_\phi$.

Verification proceeds sequentially from left to right. Let $\hat{y}_{<i}$ denote the longest verified prefix that satisfies the acceptance criterion. These tokens are committed to the final output, while the first violating token is replaced by the verifier-preferred prediction:
\begin{equation}
y^{*}_{i}
=
\arg\max_{y}
Q_{\phi}(y \mid \bm z_t, \hat{y}_{<i}),
\end{equation}
The remaining suffix tokens are discarded to avoid propagating errors from unreliable continuations.

The cloud server sends back only the verified tokens, including any corrections. The edge device then updates its decoding state to align with this verified sequence and continues generating from the last confirmed position.

\subsection{Adaptive Length $L$}
\vspace{-1mm}
The draft block length $L$ controls how frequently verification is triggered. A larger $L$ reduces cloud interactions and communication overhead, but increases the risk of error propagation before correction. In contrast, a smaller $L$ enables more frequent verification and can improve reliability, at the cost of higher communication and computation. Therefore, the optimal $L$ balances generation reliability and system efficiency.

Instead of using a fixed block length $L$, we adopt an adaptive strategy that adjusts $L$ among $\{L_{\min}, L_{\text{base}}, L_{\max}\}$ to balance verification granularity and cloud overhead. The first block uses $L_{\text{base}}$, which also serves as the default. 
If the previous block is corrected by the cloud, indicating unstable local drafting, we reduce the next block length to $L_{\min}$ to increase verification frequency and limit error accumulation. If at least two consecutive blocks are fully accepted, suggesting stable predictions, we increase $L$ to $L_{\max}$ to reduce cloud queries. Otherwise, we keep $L = L_{\text{base}}$.

\section{Experimental Setup}

\begin{table}[!t]
\centering
\caption{Language-specific prompts for speech emotion captioning.}
\label{tab:prompt_raw}
\scriptsize
\setlength{\tabcolsep}{4pt}
\renewcommand{\arraystretch}{1.15}
\begin{tabularx}{\columnwidth}{@{}p{0.22\columnwidth}X@{}}
\toprule
\textbf{Type} & \textbf{Prompt} \\
\midrule
\textbf{English Prompt} &
As an expert in the field of emotions, please focus on the acoustic information in the audio to discern clues related to the emotions of the individual. Please provide a detailed description and ultimately predict the emotional state of the individual. In the audio, respond in English only, use third person, avoid dialogue style. Example: ``In the audio, the speaker's tone is raised with rapid pacing, expressing dissatisfaction. Therefore, the speaker exhibits frustration and impatience.'' \\
\midrule
\textbf{Chinese Prompt} &
\begin{CJK*}{UTF8}{gbsn}
任务：请基于给定音频，输出一句“情感说明短句”。必须遵守：只输出一句中文短句（12--30个汉字），以“。”结尾；句子中同时包含一个主要情绪和一个简短的声学或韵律线索（如语气、语速、强弱、音高变化等类别层面的描述即可），但不要解释或列举；不要出现客套话、邀请继续对话、表情符号、英文、Markdown、标号或代码；不要提及“音频”“模型”“分析”或“我”；若存在多种可能性，只选择最可能的一种，不要并列罗列。只给出最终这一句短句，不要输出其他内容。
\end{CJK*} \\
\bottomrule
\end{tabularx}
\end{table}

\noindent \textbf{Dataset and Prompt.} We evaluate UGSD on MER2024~\cite{lian2024mer} using English and Chinese subsets, with 332 recordings for each language. Each recording is paired with one or more human-written reference captions in the corresponding language. All systems use the same prompt shown in Table~\ref{tab:prompt_raw}.%\td{the prompt shows respond in english only. Is this true for Chinese data as well? }

\noindent \textbf{Implementation Details.}
We use Qwen2.5-Omni-3B as the edge draft backbone and Qwen3-Omni-30B-A3B-Instruct as the cloud verifier backbone~\cite{qwen25omni,qwen3omni}
%\td{check ref 22 where you have qwen32.5}.
For UGSD, we optimize $L$ within the range of $[3,50]$. The adaptive $L$ is optimized as $L_{\min}=3$, $L_{\text{base}}=5$, and $L_{\max}=7$. We compare edge-only, cloud-only, and UGSD in a controlled emulation setting: the edge runs FP32 on two CPU cores, and the cloud runs bfloat16 on an NVIDIA A100 node. We report caption quality with BLEU-1, BLEU-4, METEOR, and ROUGE-L; efficiency with time to first token (TTFT), input tokens per second (ITPS), output end time (OET), output tokens per second (OTPS), and total inference time; and resource usage with CPU/RAM on the edge and GPU utilization/memory on the cloud~\cite{wang2024efficient,wang2025healthslm}. We optimized the hyperparameter of the acceptance threshold \(R\) = 20 on the cloud for all languages.
\newcommand{\method}{\textsc{UGSD}} % Uncertainty-Guided Speculative Decoding

\section{Results}
\FloatBarrier
We report the performance of our method in terms of caption quality, efficiency, and token privacy compared with edge and cloud methods. We also analyze how model size affects these results and the edge–cloud trade-off.

\subsection{Performance Comparison}

\begin{table*}[!h]
\centering
\caption{Comparison of latency and resource usage. %\td{also use your name to replace spectulative decoding. }
Lower is better for TTFT, OET, Total Time, CPU, RAM, GPU, and GPU Mem ($\downarrow$). Higher is better for ITPS and OTPS ($\uparrow$). Where bold indicates the best, and underlining indicates the second best.} 
\label{tab:latency_throughput}
\scriptsize
\setlength{\tabcolsep}{3pt}
\renewcommand{\arraystretch}{1.05}
\begin{tabular}{@{}lccccccccc@{}}
\toprule
\textbf{Config} &
\textbf{TTFT$\downarrow$} &
\textbf{ITPS$\uparrow$} &
\textbf{OET$\downarrow$} &
\textbf{OTPS$\uparrow$} &
\textbf{Total Time$\downarrow$} &
\textbf{CPU (\%)$\downarrow$} &
\textbf{RAM (GB)$\downarrow$} &
\textbf{GPU (\%)$\downarrow$} &
\textbf{GPU Mem (GB)$\downarrow$} \\
\midrule
Edge (3B) & 8.41 & 2340.11 & 40.12 & 1.53 & 40.21 & 201.6 & 20.55 & —— & —— \\\midrule
\method{} ($L=3$)  & 6.23 & 2504.27 & 2.89 & \underline{12.92} & \textbf{28.47} & 68.8 & 2.87 & 44.2 & \underline{37.75} \\
\method{} ($L=5$)  & \underline{6.18} & \textbf{2561.42} & \underline{2.73} & 12.85 & 28.73 & 68.8 & 2.83 & 44.5 & \underline{37.75} \\
\method{} ($L=7$)  & 6.30 & 2428.09 & 2.79 & 9.76  & 29.56 & \textbf{66.6} & \textbf{2.77} & 43.5 & 37.76 \\
\method{} ($L=10$) & 6.40 & 2439.40 & 2.77 & 8.60  & 28.92 & \underline{66.8} & \underline{2.78} & 43.0 & 37.76 \\
\method{} ($L=20$) & 6.75 & 2395.87 & 2.78 & 3.96  & 29.31 & 68.0 & 2.79 & \textbf{42.5} & \underline{37.75} \\
\method{} ($L=50$) & 7.58 & 2430.07 & 5.10 & 2.04  & 31.84 & 73.5 & 3.01 & \underline{43.4} & 37.82 \\
\midrule
\method{} (dynamic $L$) & \textbf{6.18} & \underline{2505.78} & \textbf{2.73} & \textbf{13.05} & \underline{28.67} & 70.4 & 2.90 & 44.9 & \textbf{37.75} \\
\bottomrule
\end{tabular}
\end{table*}

\begin{table}[!t]
\centering
\caption{Comparison of caption quality on the English subset. Spec means speculative decoding%\td{use your method name? Spec is also your method, but with fixed L, so you can replace spec with ugsd.}
.}
\label{tab:quality_only}
\scriptsize
\setlength{\tabcolsep}{5pt}
\begin{tabular}{@{}p{2.2cm}cccc@{}}
\toprule
\textbf{Model} &
\textbf{BLEU-1$\uparrow$} &
\textbf{BLEU-4$\uparrow$} &
\textbf{METEOR$\uparrow$} &
\textbf{ROUGE-L$\uparrow$} \\
\midrule
Edge          & 36.71 & 0.71 & 13.92 & 14.38 \\
Cloud         & 48.19 & 1.20 & 23.68 & 24.01 \\ \midrule
\method{} ($L=3$)  & 39.36 & \underline{1.05} & 24.15 & 19.89 \\
\method{} ($L=5$)  & 43.75 & 1.02 & \underline{24.32} & \underline{20.27} \\
\method{} ($L=7$)  & \underline{44.05} & 1.02 & 23.65 & 18.46 \\
\method{} ($L=10$) & 42.66 & 1.01 & 23.27 & 17.16 \\
\method{} ($L=20$) & 42.88 & 1.00 & 21.70 & 13.56 \\
\method{} ($L=50$) & 37.92 & 1.00 & 21.13 & 12.30 \\
\midrule
\method{} (dynamic $L$) & \textbf{44.65} & \textbf{1.06} & \textbf{24.56} & \textbf{20.69} \\
\bottomrule
\end{tabular}
\end{table}

\begin{table}[!t]
\centering
\caption{Comparison of caption quality on the Chinese subset.}
\label{tab:quality_chinese_pretty}
\scriptsize
\setlength{\tabcolsep}{5pt}
\begin{tabular}{@{}p{2.2cm}cccc@{}}
\toprule
\textbf{Method} &
\textbf{BLEU-1$\uparrow$} &
\textbf{BLEU-4$\uparrow$} &
\textbf{METEOR$\uparrow$} &
\textbf{ROUGE-L$\uparrow$} \\
\midrule
Edge          & 39.14 & 2.01 & 12.35 & 16.77 \\
Cloud         & 51.04 & 3.77 & 25.01 & 26.61 \\ \midrule
\method{} ($L=3$)  & 41.89 & \underline{3.23} & 25.62 & 22.40 \\
\method{} ($L=5$)  & 46.44 & 3.12 & \underline{25.84} & \underline{22.79} \\
\method{} ($L=7$)  & \underline{46.75} & 3.12 & 24.97 & 20.94 \\
\method{} ($L=10$) & 45.31 & 3.09 & 24.48 & 19.61 \\
\method{} ($L=20$) & 45.54 & 3.05 & 22.44 & 15.93 \\
\method{} ($L=50$) & 40.39 & 3.05 & 21.70 & 14.64 \\
\midrule
\method{} (dynamic $L$) & \textbf{47.37} & \textbf{3.27} & \textbf{26.15} & \textbf{23.22} \\
\bottomrule
\end{tabular}
\end{table}
\vspace{-1mm}

\noindent \textbf{Caption Quality.}
Tables~\ref{tab:quality_only} and~\ref{tab:quality_chinese_pretty} summarize caption quality on the English and Chinese subsets.
On the English subset (Table~\ref{tab:quality_only}), the edge-only baseline underperforms the cloud-only model due to its limited capacity.
Fixed-$L$ speculative decoding already yields substantial gains over edge-only; among the tested values, moderate block lengths perform best, and the strongest fixed setting is achieved at $L{=}7$.
\method{} with dynamic $L$ achieves the best overall performance, showing relative improvements ranging from 21.6\% to 76.4\% in terms of BLEU, ROUGE-L and METEOR. 
% improving BLEU-1 from 36.71 to 44.65 (\textbf{+21.6\%}), ROUGE-L from 14.38 to 20.69 (\textbf{+43.9\%}), and METEOR from 13.92 to 24.56 (\textbf{+76.4\%}).
Notably, \method{} closes about 69\% of the BLEU-1 gap and about 66\% of the ROUGE-L gap between edge-only and cloud-only, indicating that UGSD %verification 
substantially narrows the quality gap to full-cloud generation while retaining an edge-first workflow.

A consistent trend is observed on the Chinese subset (Table~\ref{tab:quality_chinese_pretty}).
Fixed-$L$ settings again provide clear improvements over edge-only, with the best fixed length at $L{=}7$.
\method{} with dynamic $L$ performs best overall, achieving improvements of 21.0\% in BLEU-1, 38.5\% in ROUGE-L, and 111.7\% in METEOR over edge-only baseline.
It similarly closes about 69\% of the BLEU-1 gap and about 66\% of the ROUGE-L gap to cloud-only baseline, suggesting that the benefit generalizes across languages.

\vspace{5pt}
\noindent \textbf{Efficiency.} Compared with edge-only inference, \method{}(with dynamic $L$) %\td{with dynamic $L$?} 
also significantly improves the efficiency. It reduces TTFT from 8.41\,s to 6.18\,s with 26.5\% reduction, and decreases total time from 40.21s to 28.67s with 28.7\% reduction.
Token emission is accelerated, with OET reduced from 40.12s to 2.73s (93.2\% reduction) and OTPS increasing from 1.53 to 13.05 (8.5$\times$).
Prefill throughput is maintained while ITPS improves with 7.1\% improvements.

Across fixed $L$, total time remains within a narrow range (28.13--31.84\,s), moderate lengths ($L=3$--$10$) remain consistently near-optimal in total runtime. %\td{here do you want to say the consistent one also improves?}
\method{} (dynamic $L$) stays within 2\% of the best fixed-$L$ runtime while offering the most robust caption quality (Tables~\ref{tab:quality_only} and~\ref{tab:quality_chinese_pretty}).
In terms of edge resource footprint, \method{}(with dynamic $L$) reduces CPU utilization from 201.6\% to 70.4\% (65.1\% reduction) and RAM from 20.55\,GB to 2.90\,GB (85.9\% reduction), %\td{@eric, is this for dynamic or fixed? For this subsection, also follow some order like discussing the fixed or the dynamic first and then to the other, don't mix them hard to follow. }
consistent with an edge-first design where lightweight drafting runs locally and cloud verification is invoked on demand.
%\td{use underline for the 2nd best performance in all tables.}

\subsection{Token Privacy}
Our framework keeps raw waveforms on-device and invokes cloud interaction only when predictive uncertainty exceeds the verification threshold. Instead of transmitting the full speech signal, the edge sends only compact representations, including the accepted text prefix, drafted token IDs for the current block, and compact audio features. 
We quantify this behavior using the token transformation rate $\rho$, defined as the proportion of drafted tokens transmitted to the cloud verifier for refinement as
$
\rho = \frac{N_{\text{transmitted}}}{N_{\text{total}}}
$. 
% The uncertainty threshold $\gamma$ controls the transformation behavior: a higher $\gamma$ yields a lower $\rho$ and fewer cloud interactions, while a lower $\gamma$ increases $\rho$ and verification frequency. Thus, selecting an appropriate $\gamma$ balances privacy protection and generation performance.

Under UGSD (dynamic $L$), only $\rho$=18.2\% of drafted tokens %\td{with dynamic $L$?}
are transmitted for cloud verification, indicating that most decoding remains local. This selective transmission reduces cross-boundary data transfer relative to fully cloud-based inference and supports a better privacy--utility trade-off, although the transmitted compact representations may still retain partial speaker or content information.%\td{need a more comprehensive discussion.}

\subsection{Impact of the Model Size}
\FloatBarrier
\begin{table}[!t]
\centering
\caption{Caption quality and privacy-preserving comparison between edge, cloud, and proposed method in UGSD (coll.\ denotes collaboration).}
\label{tab:quality_efficiency_main}
\scriptsize
\setlength{\tabcolsep}{4pt}
\begin{tabular}{@{}p{1.8cm}ccccc@{}}
\toprule
\textbf{Model} & \textbf{BLEU-1$\uparrow$} & \textbf{BLEU-4$\uparrow$} & \textbf{METEOR$\uparrow$} & \textbf{ROUGE-L$\uparrow$} & \textbf{Privacy} \\
\midrule
Cloud (7B, A100)     & 39.56 & 0.89 & 10.60 & 14.59 & $\times$ \\
Edge (3B, CPU)       & 36.71 & 0.71 & 13.92 & 14.38 & \checkmark \\
UGSD (coll.)   & 37.63 & 0.78 & 13.93 & 14.41 & \checkmark \\
\bottomrule
\end{tabular}
\end{table}

\vspace{-1mm}

We evaluate a smaller cloud verifier, which narrows the edge–cloud capacity gap, to test whether UGSD still maintains its effectiveness and efficiency. %reduce the performance gap between edge and cloud models. 
Specifically, we replace the cloud verifier with a 7B model while keeping the edge draft model as a 3B CPU model. As shown in Table~\ref{tab:quality_efficiency_main}, gains become smaller but remain positive: With a 7B verifier, UGSD improves BLEU-1 from 36.71 to 37.63 (+2.5\%) and ROUGE-L from 14.38 to 14.41 over edge-only. This is expected because a smaller verifier provides less reliable corrective feedback on difficult spans. However, the improvements still validates the effectiveness of UGSD regardless of the model sizes. %Importantly, the collaboration protocol is unchanged: the cloud is invoked only under elevated uncertainty and provides lightweight corrections or confirmations, suggesting that the framework does not require a very large verifier to function, although larger verifiers yield larger gains.

\section{Conclusion}
% We presented a privacy aware edge-cloud collaborative framework for speech emotion captioning based on token level speculative decoding. Raw waveforms remain on device, and a cloud verifier is queried only under high predictive uncertainty, enabling on demand refinement without continuous offloading. Experiments show substantial gains in caption quality over edge only inference while reducing end to end latency, supporting an edge first tradeoff between quality and efficiency.

% We identify key deployment tradeoffs. Gains depend on verifier capacity, with smaller cloud models yielding more modest improvements, and the draft block size controls the balance between caption reliability and system cost, motivating adaptive policies that respond to local uncertainty. Future work will study stronger uncertainty signals beyond entropy, tighter communication budgets to further reduce cross boundary transmission, and robust operation in real world mobile settings.

We presented an effective, efficient and private edge--cloud collaborative framework for SEC that leverages token-level speculative decoding. The proposed design keeps raw waveforms on device and invokes a cloud verifier only when uncertainty rises, enabling on-demand assistance while avoiding continuous offloading. Our approach substantially improves caption quality over edge-only inference and reduces end-to-end latency relative to running the entire pipeline on the device, yielding a practical quality--efficiency compromise for edge-first deployment.

% Our study also highlights several deployment-relevant trade-offs. The gains depend on verifier capacity: smaller cloud models provide only modest improvements, suggesting that corrective feedback quality is a key factor for collaboration effectiveness. In addition, the draft block size controls a clear quality--efficiency frontier, motivating adaptive policies that react to local drift. Future work will study stronger uncertainty signals beyond entropy, tighter communication budgets that further reduce cross-boundary disclosure, and robustness under unstable networks and intermittent connectivity, with the goal of supporting reliable SEC in real-world mobile settings.

%\td{references are not consistent; some have full author list and some have et al, make them consistent. Also double check your references to make sure there are no hallucinations or wrong information.}
\bibliographystyle{IEEEtran}
\bibliography{mybib}

\end{document}